\documentclass[aps,prl,twocolumn,showpacs]{revtex4-1}
\usepackage{amsmath}
\usepackage{graphicx}
\usepackage{amssymb}
\usepackage{amsthm}
\usepackage{hyperref}
\usepackage[english]{babel}
\usepackage{pgf}
\usepackage{tikz}

\newcommand{\skp}[1]{ \left \langle #1 \right \rangle }

\newcommand{\set}[1]{\mathbb{#1}}

\begin{document}

\title{Bouncing of charged droplets: An explanation using mean curvature flow}
\date{\today}
\author{Sebastian Helmensdorfer}
\email{shelmensdorfer@googlemail.com}
\affiliation{Sintzenichstr. 11, 81479 M\"unchen, Germany}
\author{Peter Topping}
\email{p.m.topping@warwick.ac.uk}
\affiliation{Mathematics Institute, Zeeman Building, University of Warwick, Coventry CV4 7AL}
\pacs{47.55.df, 47.55.nk}

\begin{abstract}
Two oppositely charged droplets of (say) water in e.g. oil or air  will tend to drift together under the influence of their charges. As they make contact, one might expect them to coalesce and form one large droplet, and this indeed happens when the charge difference is sufficiently small. However, 
Ristenpart et al discovered a remarkable physical phenomenon whereby for large enough charge differentials, the droplets bounce off each other as they make contact. Explanations based on minimisation of area under a volume constraint have been proposed 
based on the premise that consideration of surface energy cannot be sufficient. However, in this note we explain that on the contrary, the bouncing phenomenon can be completely explained in terms of energy, including an accurate prediction of the threshold charge differential between coalescence and bouncing.  
\end{abstract}

\maketitle

Droplet motion induced by electrical charges has been extensively studied at least since the 19th century (see e.g. \cite{lordrayleigh, ristenpartal2}.
Such a process occurs in a wide variety of applications such as storm cloud formation, commercial ink-jet printing, petroleum and vegetable oil dehydration, electrospray ionization for use in mass spectrometry, electrowetting and lab-on-a-chip manipulations (see e.g. \cite{trauetal, leunissenetal, baygentssaville, ochsczys, eowetal, fennetal, baretmugele, linketal, calvert}), most of which are only partially understood.

\begin{figure*}
  \caption{Coalescence and bouncing of water in oil}
  \label{fig:experiment1}
  \includegraphics[scale=0.6]{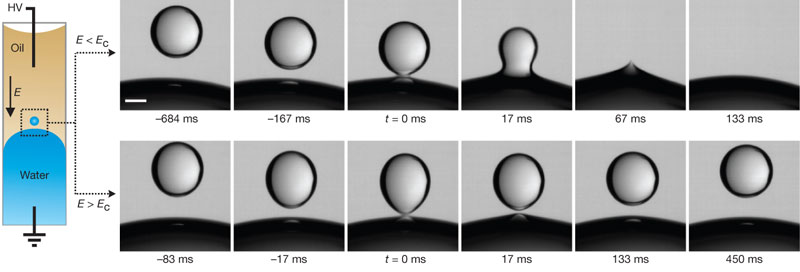} \\
  \emph{A high-voltage (HV) provides the electric field of strength $E$. The top row of images shows coalescence ($E < E_C$) whereas the bottom row shows bouncing ($E > E_C$) of the water droplet.
Reprinted by permission from Macmillan Publishers Ltd: Nature,
WD Ristenpart et al. {\bf 461}, 377-380, \copyright 2009, doi:10.1038/nature08294.  
}
\end{figure*}

In this note, we are interested in the physics of two droplets of (say) 
water, of differing charges, that move within 
a somewhat electrically insulating immiscible fluid such as 
oil. 
Under the influence of the charge difference the water droplets 
approach each other until they touch, at which point one might expect the droplets to coalesce into one droplet. While this absorption does indeed occur when the charge difference is small enough, it is a remarkable discovery of
Ristenpart et al \cite{ristenpartal1} that droplets 
with a high charge difference
tend to \emph{bounce off} each other. This is illustrated in Figure \ref{fig:experiment1}, where a water column in oil is used instead of a second droplet (the phenomenon is the same as with two droplets).
\begin{figure}
  \caption{Conical structure}
  \label{fig:experiment2}
  \includegraphics[scale=0.48]{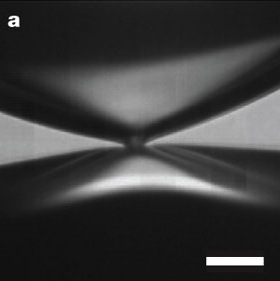} \\
  \emph{As the droplets touch, they are locally conical, and form a
thin connecting bridge. Reprinted by permission from
Macmillan Publishers Ltd: Nature, WD Ristenpart et al.
461, 377-380, 
c 2009, doi:10.1038/nature08294.
}
\end{figure}
Various explanations have been given for this phenomenon 
\cite{ristenpartal1, ristenpartal2}
starting with the premise that dynamics driven by surface tension would always lead to coalescence. However, the purpose of this note is to explain that on the contrary, the bouncing/coalescing can be explained purely by postulating that in the instant after touching, when the charge difference disappears, the droplets move solely to reduce their surface area/energy as quickly as possible. We are then able to invoke theory of the so-called \emph{mean curvature flow} as we explain below.

\begin{figure}[h]
  
  \caption{Conical structure 2}
  \label{fig:doubleconeangle}
  \begin{tikzpicture}[scale=2]
  \draw [thin, dashed, gray, ->] (0,-1.65) -- (0,1.65);
  \draw [thin, dashed, gray, ->] (-1,0) -- (1,0);

  \draw  (1,-1.5) -- (0.3,-0.45);
  \draw  (0.3,0.45) -- (1,1.5);
  \draw  (-1,-1.5) -- (-0.3,-0.45);
  \draw  (-0.3,0.45) -- (-1,1.5);
  \draw [dotted] (0,-1.5) ellipse (1cm and 0.1cm);
  \draw [dotted] (0,1.5) ellipse (1cm and 0.1cm);

  \draw  (0.3,-0.45) .. controls (0.03,0) .. (0.3,0.45);
  \draw  (-0.3,-0.45) .. controls (-0.03,0) .. (-0.3,0.45);

  \draw [<->, thin] (0.8,0)  arc (0:56.31:0.8) node [right=7pt, below=12pt] {$\theta$};
\end{tikzpicture} \\
 \emph{The double cone angle $\theta$ is proportional to the field strength and determines the behaviour of the system.}
\end{figure}
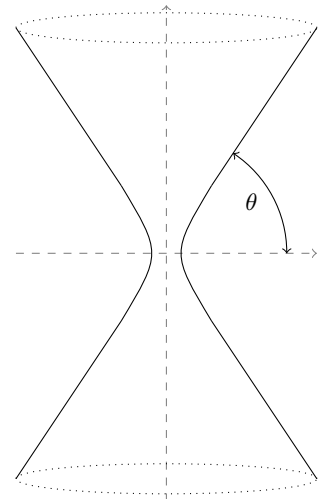

\begin{figure}[ht]
  \caption{Double cone angle and potential}
  \label{fig:experiment3}
  \includegraphics[scale=0.4]{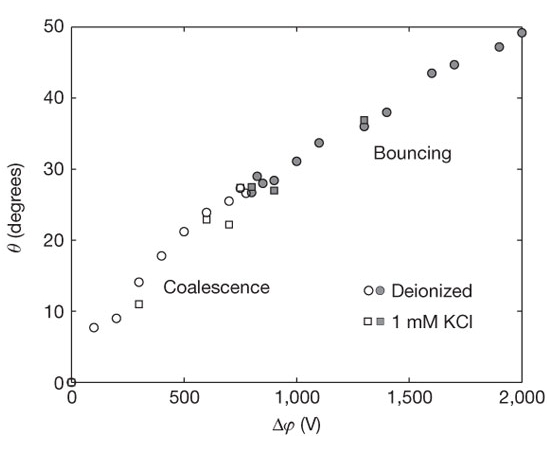} \\
  \emph{The contact angle $\theta$ of the double cone for water droplets in air with applied potential $\triangle \varphi$. Open symbols denote coalescence, filled symbols denote bouncing. 
Reprinted by permission from Macmillan Publishers Ltd: Nature,
WD Ristenpart et al. {\bf 461}, 377-380, \copyright 2009, doi:10.1038/nature08294.  
}
\end{figure}

Our starting point is the well-understood principle that as the droplets approach each other, they deform into a conical shape (cf. \emph{Taylor cones} \cite{taylor, saville, stoneetal, oddershede, collinsetal, melcher, delamora}) under the influence of the differing charges, and as they touch, the droplets will look locally like a double cone with a thin connecting bridge, as illustrated in Figures \ref{fig:experiment2} and \ref{fig:doubleconeangle}.
Experiment shows (see Figure \ref{fig:experiment3})
that the cone angle $\theta$ in the regime we are considering 
is proportional to the charge difference, and that there is a critical threshold angle $\theta_C\simeq 27^\circ$ below which one observes coalescence, and above which one observes bouncing, independently of the choice of fluids.

Our goal in this note is to survey the theory of \emph{mean curvature flow from cones}, which governs how a cone can evolve in order to reduce its surface area as quickly as possible, and show that the behaviour of this flow undergoes a bifurcation at a critical cone angle of $24^\circ$.
We then indicate how abstract arguments prove rigorously that evolutions of the touching droplets \emph{must} show coalescence for cones angles below the critical angle, and \emph{must} show bouncing for angles above the critical angle.

It is a feature of our approach that we need make no symmetry or self-similarity assumptions on the evolution of the fluid
(although self-similar solutions feature heavily in our analysis). Our model requires few physical assumptions and there are no parameters to fit.
We know from \cite{ristenpartal2}
that the charges are equalised at the moment of touching, and thus 
we postulate that in the instant after touching, the droplets evolve in order to reduce their surface area as quickly as possible.
However, this evolution is occurring only \emph{locally} in time and space, so it is irrelevant that global minimisation of surface area would always result in coalescence. Ultimately, all the subtlety of the distinction between coalescence and bouncing arises from the mathematical theory of mean curvature flow.



Consider a one-parameter family of surfaces $M_t$ in 
$\set{R}^3$ that can be viewed as the images of immersions
$F_t:M\to\set{R}^3$ from a fixed surface $M$.
When the area $A_t$ of these surfaces is finite, it evolves by
\begin{equation}
\frac{ d }{ d t} A_t = - \int_{M_t} \skp{ \vec{H}, \frac{ \partial }{\partial t } F_t } d\mu_{M_t}
\end{equation}
where the normal vector field $\vec{H}$ on the surface is called the mean curvature (see 
\cite[Appendix A]{eckermcf}
for a direct definition).
The evolution of the surface that reduces the area as quickly as possible (more precisely the $L^2$-gradient flow) is then the much-studied mean curvature flow (MCF) \cite{eckermcf} defined by the nonlinear PDE
\begin{equation}
  \label{eq:mcfequation}
  \frac{ \partial }{ \partial t } F_t = \vec{H}.
\end{equation}
The comparison principle \cite{eckermcf} tells us that two different solutions that are disjoint initially at time $t=0$, will remain disjoint for later times $t>0$.

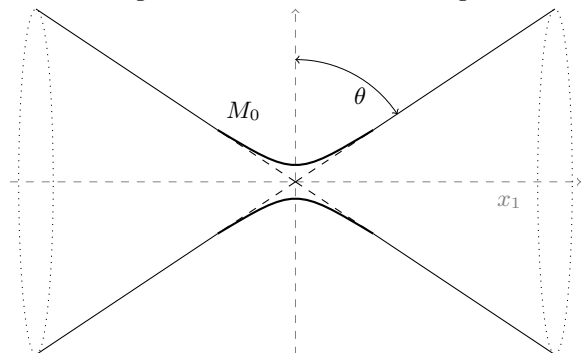
\begin{figure}[h]
  
  \caption{Double cone smoothings}
  \label{fig:theory1}
  \begin{tikzpicture}[scale=2.3]

  \draw [very thin, dashed, gray, ->] (0,-1) --  (0,1);
  \draw [very thin, dashed, gray, ->] (-1.65,0) -- node[very near end, below = 2pt] {$x_1$} (1.65,0);

  \draw  (-1.5,1) -- (-0.45,0.3);
  \draw  (0.45,0.3) --  (1.5,1);
  \draw  (-1.5,-1) -- (-0.45,-0.3);
  \draw  (0.45,-0.3) -- (1.5,-1);
  \draw [dashed] (0,0) -- (-0.45,0.3);
  \draw [dashed] (0,0) -- (0.45,0.3);
  \draw [dashed] (0,0) -- (0.45,-0.3);
  \draw [dashed] (0,0) -- (-0.45,-0.3);
  \draw [dotted] (-1.5,0) ellipse (0.1cm and 1cm);
  \draw [dotted] (1.5,0) ellipse (0.1cm and 1cm);

  \draw [thick] (-0.45,0.3) .. controls (0,0.03) .. node[very near start, above=6pt, fill=white] {$M_0$} (0.45,0.3);
  \draw [thick] (-0.45,-0.3) .. controls (0,-0.03) .. (0.45,-0.3);

  \path (0,0.5) -- node[near end, fill=white] {$\theta$} (0.5,0.5);
  \draw [very thin, <->] (0,0.7071) arc (90:33.69:0.7071cm);

\end{tikzpicture} \\

\emph{We smooth out the double cone $M_0^\theta$ in order to obtain a rotationally symmetric surface $M_0$, locally modeling the touching fluid droplets.}

\end{figure}

We locally model the touching fluid droplets as follows (Figure \ref{fig:theory1} shows the construction). 
We rotate the function $u_\theta(x_1):=|x_1|/\tan\theta$ 
around the $x_1$ axis in $\set{R}^3$ to get a double cone 
$M^\theta_0$, and model the fluid bridge with an arbitrary one-sheeted 
surface $M_0$ that smooths out $M^\theta_0$
as in Figure \ref{fig:theory1}.


We can solve the mean curvature flow starting with this $M_0$, and it will either flow for all time (corresponding to coalescence) or will develop a singularity at finite time as the thin neck shrinks to nothing (meaning the bridge between the droplets breaks apart and we get bouncing) \cite{helmensdorfer}.

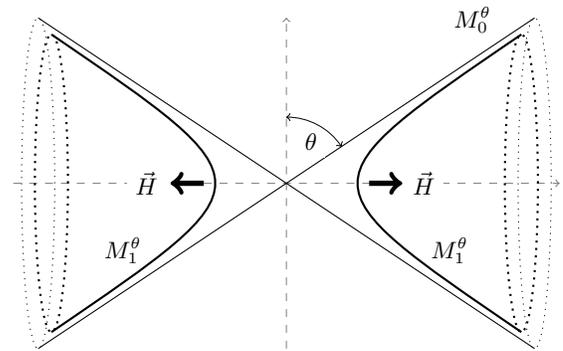
\begin{figure}[h]
  \caption{One-sheeted and two-sheeted evolutions of double cones}
  \label{fig:experiment4}

  \begin{tikzpicture}[scale=2.2]
  \draw [very thin, dashed, gray, ->] (0,-1.0) -- (0,1.0);
  \draw [very thin, dashed, gray, ->] (-1.65,0) -- (1.65,0);

  \draw [thin]  (-1.5,1) -- (0,0) -- node [near end, above=7pt, fill=white] {$M_0^\theta$} (1.5,1);
  \draw [thin] (-1.5,-1) -- (0,0) -- (1.5,-1);
  \draw [thin, dotted] (-1.5,0) ellipse (0.1cm and 1cm);
  \draw [thin, dotted] (1.5,0) ellipse (0.1cm and 1cm);

  \draw [thick] (-1.42,0.9) .. controls (-0.1,0) .. node[very near end, above=4pt] {$M_1^\theta$} (-1.42,-0.9);
  \draw [thick] (1.42,0.9) .. controls (0.1,0) .. node[very near end, above=4pt] {$M_1^\theta$} (1.42,-0.9);
  \draw [thick, dotted] (-1.42,0) ellipse (0.1cm and 0.9cm);
  \draw [thick, dotted] (1.42,0) ellipse (0.1cm and 0.9cm);

  \path (0,0) -- node[very near end, left=7pt, fill=white] {$\theta$} (0.4,0.29);
  \draw [very thin, <->] (0,0.4) arc (90:33.69:0.4cm);

  \draw [line width=2pt, ->] (0.5, 0) -- node[near end, right=3pt, fill=white] {$\vec{H}$} (0.7,0);
  \draw [line width=2pt, ->] (-0.5, 0) -- node[near end, left=4pt, fill=white] {$\vec{H}$} (-0.7,0);

\end{tikzpicture} \\

a. \emph{The two-sheeted evolution of $M^\theta_0$ exists for any angle $\theta$. It is the unique evolution of $M_0^\theta$ for $\theta > \theta_C^*$ and corresponds to the droplets bouncing off each other.} \\
$ $ \\

 \begin{tikzpicture}[scale=2.2]

  \draw [very thin, dashed, gray, ->] (0,-1) -- (0,1);
  \draw [very thin, dashed, gray, ->] (-1.65,0) -- (1.65,0);

  \draw [thin]  (-1.5,1) -- (0,0) -- node [near end, below=7pt, fill=white] {$M_0^\theta$} (1.5,1);
  \draw [thin] (-1.5,-1) -- (0,0) -- (1.5,-1);
  \draw [thin, dotted] (-1.5,0) ellipse (0.1cm and 1cm);
  \draw [thin, dotted] (1.5,0) ellipse (0.1cm and 1cm);

  \draw [thick] (-1.42,0.98) .. controls (0,0.15) .. (1.42,0.98);
  \draw [thick] (-1.42,-0.98) .. controls (0,-0.15) .. node[very near end, below=4pt]{$M_1^\theta$} (1.42,-0.98);
  \draw [thick, dotted] (-1.42,0) ellipse (0.1cm and 0.98cm);
  \draw [thick, dotted] (1.42,0) ellipse (0.1cm and 0.98cm);

  \path (0,0) -- node[very near end, left=7pt, fill=white] {$\theta$} (0.4,0.29);
  \draw [very thin, <->] (0,0.4) arc (90:33.69:0.4cm);

  \draw [line width=2pt, ->] (0, 0.4) -- node[near end, right=4pt, fill=white] {$\vec{H}$} (0,0.6);
  \draw [line width=2pt, ->] (0, -0.4) -- node[near end, right=4pt, fill=white] {$\vec{H}$} (0,-0.6);

\end{tikzpicture} \\

b. \emph{For $\theta < \theta_C^*$ at least two one-sheeted evolutions of $M^\theta_0$ exist. These can be used as pinching barriers and therefore imply coalescence of the droplets.}

\end{figure}

As a tool to understand which of these occurs for each $\theta$, we consider \emph{self-similar} solutions of the mean curvature flow starting at the double cone of angle $\theta$, i.e. 
solutions $M_t$ of the form $\sqrt{t}M^\theta_1$ where 
$M^\theta_1$ asymptotically approaches the double cone $M^\theta_0$.
In general, there will be many solutions of this form with the same cone as the initial surface, but if we impose the ansatz that the solution should be symmetric under rotations about the $x_1$-axis, and under reflections $x_1\mapsto -x_1$, then all solutions can be classified up to the solution of a simple ODE \cite{ilmanen, helmensdorfer}.
It turns out that for $\theta$ smaller than some critical value, which can be computed numerically \cite{ilmanen} to be 
$\theta^*_C\sim 23.96^\circ$, there are at least two such solutions that are one-sheeted (see Figure \ref{fig:experiment4}b), but as $\theta$ increases to $\theta^*_C$, these two solutions merge into one, a bifurcation occurs, and for 
$\theta>\theta^*_C$, there is \emph{no} such one-sheeted solution, only a two-sheeted solution which corresponds to bouncing (see Figure \ref{fig:experiment4}a).


Meanwhile, a simple application of a suitable comparison principle (see \cite{barlescomparison}) \emph{proves} that for cone angles less than $\theta^*_C$, \emph{any} solution of MCF starting at the smoothed cone $M_0$ \emph{must} evolve as a one-sheeted solution
(not necessarily rotationally symmetric or self-similar)
that corresponds to coalescence (see Figure 
\ref{fig:experiment4}b.).
A more subtle argument \cite{ilmanen, helmensdorfer} tells us that for cone angles greater than $\theta^*_C$, any solution of MCF starting at $M^\theta_0$ \emph{must} evolve as a two-sheeted solution (again not necessarily rotationally symmetric or self-similar)
that corresponds to bouncing (see Figure \ref{fig:experiment4}a.).  For this latter case, one imagines an envelope of all possible solutions emanating from the cone, whether rotationally symmetric or not, and argues that this envelope itself must be a new, regular solution
that is now necessarily rotationally symmetric. If there exists any non-bouncing solution starting at a cone, then this envelope solution must be a one-sheeted solution of the type classified above, which does not exist for cone angles greater than $\theta^*_C$ \cite{ilmanen, helmensdorfer}.

Our theory therefore predicts a transition between coalescence and bouncing at a critical angle which is within a few degrees of the experimentally observed value.

Finally we want to compare our approach to the one from \cite{ristenpartal1, ristenpartal2}. 
In those papers, assumptions are made about the shape of the joining neck for each cone angle, and then the sign of the mean curvature is computed to determine whether that neck should expand or pinch. With the assumptions of \cite{ristenpartal1} this leads to a prediction of $\theta_C = 45^\circ$.
In \cite{ristenpartal2} the joining neck is assumed to be of constant mean curvature locally, and connected to the linear double cone in a non-differentiable way at a radius $r_0$. 
There remains one degree of freedom in their model, and tuning this leads to different predictions of $\theta_C$.
In \cite{ristenpartal2} this free parameter
is removed by assuming that the volume of fluid 
within the ball of radius $r_0$ remains constant as the neck starts evolving, i.e. no fluid is allowed to flow in from further out, and this then 
leads to a prediction of $\theta_C = 30.8^\circ$. 

Although we make no assumption in our model that the evolution of the droplets is locally self-similar, a stability argument suggests that this is likely. In this case, our model makes a local prediction for the shape of the neck as it evolves, and in particular suggests that the radius of the neck should grow like $\sqrt t$ (in contrast to the discussion in \cite{ristenpartal2}).

\emph{Conclusion:} 
We propose a new model to explain the remarkable phenomenon of bouncing and coalescence of charged fluid droplets. Our model assumes that the system after touching moves according to the mean curvature flow, which means its area decreases as fast as possible. Analyzing the flow in this setting leads to a prediction of about $24^\circ$ for the critical angle, which is in good agreement with experiments. Therefore minimization of energy can, contrary to general belief, explain the phenomenon. 
One advantage of our approach compared to existing ones is that we do not make strong assumptions on the precise shape of the bridge between the touching fluid droplets. 


\emph{Acknowledgements:}
Both authors were supported by The Leverhulme Trust.
The second author was supported by EPSRC grant EP/K00865X/1.

\end{document}